\documentclass[conference]{IEEEtran}

\usepackage{graphicx}
\usepackage{bm}
\usepackage{cite}
\usepackage{amsmath}
\usepackage{footnote}
\usepackage{color}
\usepackage{eurosym}
\usepackage{url}

\def\apgt{\ {\raise-.5ex\hbox{$\buildrel>\over\sim$}}\ }
\def\aplt{\ {\raise-.5ex\hbox{$\buildrel<\over\sim$}}\ }
\def\lteq{\ {\raise-.5ex\hbox{$\buildrel<\over-$}}\ }

\def\JB#1{{}}

\begin{document}

\title
{24.77 Pflops on a Gravitational Tree-Code to \\Simulate the Milky Way Galaxy with 18600 GPUs}
\author{\IEEEauthorblockN{Jeroen B\'edorf}
\IEEEauthorblockA{Leiden Observatory\\
Email: bedorf@strw.leidenuniv.nl}
\and
\IEEEauthorblockN{Evghenii Gaburov}
\IEEEauthorblockA{SURFsara Amsterdam\\
Email: evghenii.gaburov@surfsara.nl}
\and
\IEEEauthorblockN{Michiko S.~Fujii}
\IEEEauthorblockA{National Astronomical Observatory of Japan\\
Email: michiko.fujii@nao.ac.jp}
\and
\IEEEauthorblockN{Keigo Nitadori}
\IEEEauthorblockA{RIKEN AICS\\
Email: keigo@riken.jp}
\and
\IEEEauthorblockN{Tomoaki Ishiyama}
\IEEEauthorblockA{CCS, University of Tsukuba\\
Email: ishiyama@ccs.tsukuba.ac.jp}
\and
\IEEEauthorblockN{Simon Portegies Zwart}
\IEEEauthorblockA{Leiden Observatory\\
Email: spz@strw.leidenuniv.nl}
}


\IEEEoverridecommandlockouts


\IEEEpubid{\begin{minipage}{\textwidth}\ \\ [12pt]
SC14, November 16-21, 2014, New Orleans\\
978-1-4799-5500-8/14/\$31.00~\copyright~2014 IEEE
\end{minipage}} 


\date{1 May 2014}
\maketitle

\urlstyle{tt}

\begin{abstract} 

We have simulated, for the first time, the long term evolution of the Milky Way
Galaxy using 51 billion particles on the Swiss Piz Daint supercomputer with our 
$N$-body gravitational tree-code {\tt Bonsai}. Herein, we describe the scientific
motivation and numerical algorithms. The Milky Way model was simulated for 6
billion years, during which the bar structure and spiral arms were fully
formed. This improves upon previous simulations by using 1000 times more
particles, and provides a wealth of new data that can be directly compared with
observations. We also report the scalability on both the Swiss Piz
Daint and the US ORNL Titan. On Piz Daint the parallel efficiency of {\tt
Bonsai} was above 95\%. The highest performance was achieved with a 242
billion particle Milky Way model using 18600 GPUs on Titan, thereby reaching a
sustained GPU and application performance of 33.49 Pflops and 24.77 Pflops
respectively.

\end{abstract}

Submitted in the categories: Scalability, Time-to-solution and Peak performance

\section{Introduction}
\label{Sect:1}

The Sun, although the bright centre of the Solar System, is only a tiny 
speck among the billions of stars that form the disk of our 
home galaxy, the Milky Way.
And, although bright and clearly visible on the night sky, 
the stars, interstellar gas and planets contain only a minor fraction of
the mass of the Galaxy as most of its mass ($\apgt 90$\,\%) is dark matter.
The stellar
distribution of the Galaxy is not smooth; the disk is warped, shows
spiral patterns, it is thicker in some places, and it is clumpy in others,
whereas the central portion of the Milky Way contains a
bar-shaped structure composed of stars.  
The origin of these structures is not
known. They may be the remnants from the formation of the Galaxy
in its cosmological setting, or such structures may appear
spontaneously, possibly due to the interaction between gas and stars
in the star-formation cycle, self-gravity, gravitational relaxation processes or 
nonlinearities in the finite point-mass dynamics of the Galaxy.

We currently lack the theoretical understanding to fully appreciate
the fine-structure dynamics of the Galaxy, and this gap between theory
and observations is only about to widen.  On 19 December 2013 the
European Space Agency launched the Gaia satellite. In 2017, this \euro1.5 billion
spacecraft will deliver its first catalogue of
about 1 billion stars throughout the Galaxy providing measurements of their distances
(up to $\sim$28,000 light-year), space velocities and stellar type.
This data will provide us with the properties of the small scale structure 
of the Milky Way disk.
By comparing these observations with simulation models, we can
recover the global structure of the Milky Way disk, including 
the pattern speed and resonances of both the bar and spiral arms.
However, it requires self-consistent simulations of the Galactic disk 
with at least an order of magnitude more stars than the number of 
Gaia samples in order to make a reliable comparison.
Such high resolution simulations capture the small scale disk structures, whereas 
in lower resolution simulations the limited number of stars washes 
out these details of the Galactic disk.
This makes it unclear how structures found in these simulations 
relate to the data that will be gathered by the Gaia satellite. 
The ultimate goal is to simulate the Galaxy on a star-by-star basis,
with 100 billion stars. This one-to-one correspondence 
between the stars in the Galaxy and the simulation will allow direct comparison
with the Gaia catalogue.

The largest $N$-body simulations of the Galaxy today contain about 
100 million objects\cite{2008NJPh...10l5002D,2009ApJ...697..293D,
2013ApJ...766...34D,2014ApJ...785..137S}.
These simulations can be categorized in two types, 
1) an analytic, static potential dark matter halo and a live ($N$-body) 
disk\cite{2008NJPh...10l5002D,2009ApJ...697..293D} or
2) both a live dark matter halo and a live disk\cite{2013ApJ...766...34D, 2014ApJ...785..137S}.
The former type of simulation allows us to accurately resolve the disk, 
but it does not capture the disk-to-halo interaction.
It is however known that angular momentum transfer from disk to halo plays 
an important role in the formation and evolution of the bar\cite{2009ApJ...697..293D}. 
For the latter simulations it requires at least five times more dark matter than disk particles
to accurately model the dark matter halo and its interaction with the disk.
As a result we have to increase the total number of particles in the simulations by almost three orders of 
magnitude, compared to previous work\cite{2009ApJ...697..293D, 2014arXiv1404.0304D},
in order to reach the desired resolution of the Galactic disk.

In this paper we present the first results of a simulation of the Milky Way
Galaxy with 51 billion particles. It is only for lack of computing
time that we have not simulated the Galaxy with 242 billion particles,
but we do present timing results and feasibility analysis.

\subsection{Simulating the Galaxy on a star-by-star basis}

By the nature of the gravitational force  each star in the Galaxy attracts all the other 100 billion stars.
Numerically such gravitational interactions are most accurately
simulated via direct $N$-body force evaluations, in which each of the
$N$ objects exerts $N-1$ interactions per time-step
\cite{2003gmbp.book.....H}. With the ${\cal O}(N^2)$ force complexity of this
algorithm, the calculation of the forces between all stars requires
$\sim10^{23}$ floating point operations. Calculating
forces is by far the most expensive operation in the Galaxy simulation;
even adding the nuclear evolution of all the stars only adds 
$\sim10^{13}$ operations and moving the stars requires 
$\sim10^{12}$ operations \cite{2013ASPC..470..353P}.

To reduce the computational cost of calculating the forces between all
the stars, we adopt the hierarchical Barnes-Hut tree algorithm
\cite{1986Natur.324..446B}.  In this algorithm the distribution of
particles is recursively partitioned into octants until the number of
particles in an octant is smaller than a critical value (we use
16~\cite{2012JCoPh.231.2825B}). Once a tree-structure is built and its multipole moments are
computed, the code proceeds with the force calculation step. For this,
we adopt a multipole acceptance criterion parameterized by an opening angle
$\theta$\cite{2012JCoPh.231.2825B}
whose purpose is to decide whether or
not the substructure in distant octants can be used as a whole.  If
the opening angle is infinitesimal the tree-code reduces to a rather
inefficient direct $N$-body code, otherwise the asymptotic complexity
of the gravitational force calculation reduces to ${\cal O}(N\log{N})$. 
With such a tree algorithm, it costs $\sim10^{16}$ floating point operations to
calculate all inter-particle forces, although some additional overhead is
introduced by building and traversing the tree-structure. 

Our choice of the Barnes-Hut tree algorithm, over a Tree Particle Mesh method
(TreePM)\cite{Ishiyama:2012:PAN:2388996.2389003,1981csup.book.....H}, to
simulate the Milky Way Galaxy is motivated by two reasons. First, the TreePM
algorithm assumes periodic boundary conditions, which makes it computationally
efficient for cosmological simulations. However, to simulate the Milky Way
Galaxy we require open boundary conditions which are computationally expensive
to use in a TreePM method, thereby defeating its main speed advantage over the
Barnes-Hut tree algorithm.  Second, the relative accuracy requirement for the
gravitational force approximations, imposed by the long-term simulation of the
Milky Way, would require a disproportionally large number of grid cells in a
TreePM method, which, due to insufficient interconnect bandwidth, would likely
cause the parallel FFT operation to become a bottleneck.

The calculation of gravitational forces is by definition an all-to-all
operation, and the tree algorithm does not lift the requirement of the
global communication in the gravitational $N$-body problem. Even though the
tree algorithm considerably reduces the number of interactions, it 
still requires all processes to communicate with each other, therefore making
this a non-trivial task to run such a code efficiently on a distributed
memory architecture with many layers of communication that have low-bandwidth
and long-latency relative to the raw performance of the compute node.

Here we describe how we solved the communication bottlenecks and how we
parallelized the global data-aware tree algorithm to make it suitable for a
large GPU-equipped supercomputer. 
Our ultimate aim is to simulate the Galaxy on a star-by-star basis. In this
work we report on our simulation of the Milky Way galaxy
with an unprecedented high resolution of 51 billion particles on the Piz Daint
supercomputer. We further demonstrate that our ultimate aim, subject to
computing time availability, can be achieved on the Titan supercomputer and will
take about a week to complete.
We used two computers for our research, Piz Daint at ETH Switzerland and Titan
at ORNL US. The time for these simulations was offered to us by
Director's Discretion.

We used 4096 GPU-equipped nodes of Piz Daint to simulate the Milky Way Galaxy
for 6 billion years with 51 billion particles. This high accuracy
simulation improves upon current state-of-the-art simulations by over 1000 times in
particle numbers and over 10 times in spatial resolution, allowing us for the first
time to resolve scales that are less than the average distance between the
stars in the Solar Neighborhood.

We also conducted an application scalability study on both Piz Daint and Titan
using up to 18600 nodes and up to a 242 billion particle Milky Way model. At
the pinnacle, we achieved a sustained GPU and application single precision
performance of 33.49\,Pflops and 24.77\,Pflops respectively.
(see \S\,\ref{Sect:Performance}). Each full step with 242 billion particles takes
about 5 seconds of wall-clock time to compute (see
Tab.\,\ref{Tab:timeBreakdown}), and with a time step of $75,000$\,year, it would
take about one week to simulate the entire Milky Way Galaxy on a star-by-star
basis for at least 8 billion years (8\,Gyr).

\section{Quantitative discussion of \\current state of the art}
\label{Sect:2}

Tree-codes have been used for previous simulations of galaxies 
with five hundred thousand \cite{2010ApJ...721L..97B} to a hundred million
\cite{2013MNRAS.431..767B,2009ApJ...703.2068D,2013ApJ...766...34D} 
particles. Each particle in these earlier simulations represents 
$\apgt 2000$ Solar masses ($M_{\odot}$). These studies are used to understand 
the structure of the bar \cite{2012MNRAS.426L..46A,
2009ApJ...697..293D,2014arXiv1404.0304D}, formation and dynamics of spiral arms
\cite{2011ApJ...730..109F,2013ApJ...763...46B,2013ApJ...766...34D},
pitch angle and galactic shear\cite{2013ApJ...763...46B,2013A&A...553A..77G}, and the
warping of the stellar disk \cite{2009ApJ...697..293D}.  
In another recent work the properties of the Milky Way bulge 
are investigated using simulations with up to 30 million 
particles~\cite{2014arXiv1404.0304D}. The largest
simulations of disk galaxies with a live halo so far have been 
performed for educational purposes~\cite{2008NJPh...10l5002D}
and for the investigation of the formation and propagation 
of spiral structure using an analytic static potential for the 
halo~\cite{2013ApJ...766...34D}. Both 
of these simulations used $\sim100$\,million particles.
Recently Fujii et al. \cite{2011ApJ...730..109F} and Sellwood
\cite{2013ApJ...769L..24S} demonstrated that the
number of particles adopted in those simulations is not sufficient to
accurately resolve relaxation processes and tend to overproduce the
dynamical heating of the disk.
Although they conclude that at least several hundred thousand particles
are required to prevent numerical heating, it is still unclear 
which resolution is required in order to accurately reproduce the intricate 
non-linear dynamics of the internal structure of a Milky Way sized galaxy. 
Furthermore, recent observation campaigns such as Gaia require even more 
accurate simulations with at least 100 times more particles to allow a 
direct comparison between observations and simulations.  
This realization has led to the requirement for improving the parallel 
performance of simulation codes before it becomes possible to simulate the
Milky Way Galaxy with sufficient resolution to resolve these issues.  
With the code described in this work, we can simulate the Galaxy on a
star-by-star basis, which is a sufficiently high resolution to address these
questions, and may therefore change our understanding of the dynamical
evolution of disk galaxies like the Milky Way.

Tree-codes have 
been used in various incarnations such as TreePM
\cite{1995ApJS...98..355X,2000ApJS..128..561B, 2002JApA...23..185B,
  2004NewA....9..111D, 2005MNRAS.364.1105S,2005PASJ...57..849Y,
  2009PASJ...61.1319I,2010IEEEC..43...63P} for previous Gordon Bell
Prize simulations \cite{Warren:1992:ANS:147877.148090,
  Warren:1998:AAC:509058.509130, 
  Kawai:1999:ANB:331532.331598, Hamada:2009:THN:1654059.1654123,
  Hamada:2010:TAN:1884643.1884644, Ishiyama:2012:PAN:2388996.2389003}.
Recently, Ishiyama et al. \cite{Ishiyama:2012:PAN:2388996.2389003} 
achieved 4.45 Pflops average performance on 82944 nodes (663552 CPU cores) 
of K computer with a cosmological tree particle-mesh calculation using 
one trillion particles.
Most of the previous tree and TreePM codes have been tuned for massively
parallel supercomputers without accelerators
\cite{1991BAAS...23.1345W,2004NewA....9..111D, 2005MNRAS.364.1105S,
  2009PASJ...61.1319I, 2010IEEEC..43...63P,
  Ishiyama:2012:PAN:2388996.2389003} except for codes optimized for
relatively small clusters including accelerators such as GPUs and
GRAPEs \cite{2004PASJ...56..521M, 2005PASJ...57..849Y,
  Hamada:2009:THN:1654059.1654123,
  Hamada:2010:TAN:1884643.1884644,2012arXiv1206.1199N}.  
The first calculation with a parallel tree-code with GPUs was running
a 1.6 billion particle cosmological dark-matter simulation on 256 GPUs
\cite{Hamada:2009:THN:1654059.1654123}. They achieved a sustained
performance of 42 Tflops and won the Gordon Bell Prize with their
performance/price ratio of 124 Mflops/\$.  The next year the same team
\cite{Hamada:2010:TAN:1884643.1884644} received an honorable mention
of the Gordon Bell Prize committee for their 190 Tflops cosmological
simulation with 3.3 billion particles on 576 GPUs on the DEGIMA
supercomputer (at the Nagasaki Advanced Computing Center).  Their
prize performance ratio was 254.4 Mflops/\$. They used the GPUs as
gravity accelerators, in very much the same way as the GRAPE
\cite{1990Natur.345...33S, 1998sssp.book.....M} family of special
purpose computers and the first GPUs
\cite{Nyland04,2007NewA...12..641P} have been used
\cite{2004PASJ...56..521M}.  In these calculations the host processors
were used for constructing the tree and walking the tree-structure to
build interaction lists, and the GPU was used for the force
calculations only. In their calculations constructing and walking the
tree (${\cal O}(N \log{N}$) operations) becomes a bottleneck, as does
the communication with the GPU.

The move from CPU-based to GPU-based supercomputers is
motivated by lower energy consumption per flop for the latter. 
For example, K computer offers 830 Mflops/watt
compared to 2.1 (2.7) Gflops/watt for
Titan (Piz Daint) \footnote{see {\tt http://www.green500.org/}}. 
We therefore expect that new exascale supercomputers will be equipped 
with similar type of accelerators (see \S\,\ref{Sect:7}).
However, on Titan the theoretical peak performance is 3.95
Tflops per node (in single precision and excluding CPU performance),
wheareas it is 0.128 Tflops per node on K computer.  The memory and network
performance per flop is therefore much lower on Titan than it is on
K computer.  While achieving high performance on the GPU, we increase
the need for a high-speed network, which requires an even more
aggressive strategy for minimizing the communication cost.

Designing a strategy to minimize the CPU-GPU and network communication will
therefore result in a general improvement of a wide variety of
algorithms because our optimization strategy is not specific to
$N$-body simulations. In our solution we reduced this 
communication by having the GPU perform the tree construction
and tree-walk, as well as the force calculations. The communication
between nodes is minimized by using a careful selection strategy of which data 
we send as well as by reusing this data for multiple purposes.  
The resulting implementation is not specific to
Titan or Piz Daint, but, in fact, can be applied on any large GPU-equipped
supercomputer, such as the TSUBAME (at the Tokyo institute of
Technology's Global Scientific information Center) series of
supercomputers, HA-PACS (at University of Tsukuba), Tianhe-1A (National
Supercomputing Center of Tianjin), Nebulae (NSCS), PLX-GPU (CINECA)
and LGM (Leiden). 

The two important improvements compared
to earlier work with parallel tree-codes, are: 1) porting the
tree-code, including the tree-building and tree-walk, entirely to the
GPU and 2) utilizing the node CPU for orchestrating the communication,
administrative purposes, feeding the GPU, and data
reduction. Improvement 1 allowed us to achieve high performance by
efficiently utilizing each GPU and improvement 2 enabled us to scale
the code to 18600 nodes even with a rather modest number of 
particles compared to previous GBP winners.

\section{What the innovations are and \\how they were achieved}
\label{Sect:3}

Given the enormous computational capabilities of the NVIDIA Kepler GPU,
communication between the MPI processes can easily become the major
scalability barrier, especially when there are $1000+$
nodes\footnote{We use one GPU per MPI process.}.  The code used in
this work is our parallel GPU tree code called {\tt Bonsai}\footnote{
  {\tt Bonsai} is publicly available as part of the AMUSE 
framework\cite{2013CoPhC.184..456P} at  \url{http://amusecode.org} or as standalone 
  tool at \url{https://github.com/treecode/Bonsai}.}.  Here we
shortly describe single GPU optimizations for the Kepler architecture
and cover in detail the parallelization scheme used to distribute the
work across multiple GPUs. For more details about the original single
GPU version we refer the reader to our previous
work~\cite{2012JCoPh.231.2825B}.

\subsection{Single GPU Optimizations}\label{Sect:GPUoptimizations}

The single GPU code consists of three major parts: tree-construction, computation of
multipole moments, and tree-walk in which inter-particle gravitational forces
are computed. In contrast to previous works
\cite{Hamada:2009:THN:1654059.1654123, Hamada:2010:TAN:1884643.1884644}, in
{\tt Bonsai} all of these steps are carried out on the GPU, leaving the CPU
with lightweight tasks such as data management and kernel launches.

\begin{figure}[t]
\begin{center}
\includegraphics[width=90mm]{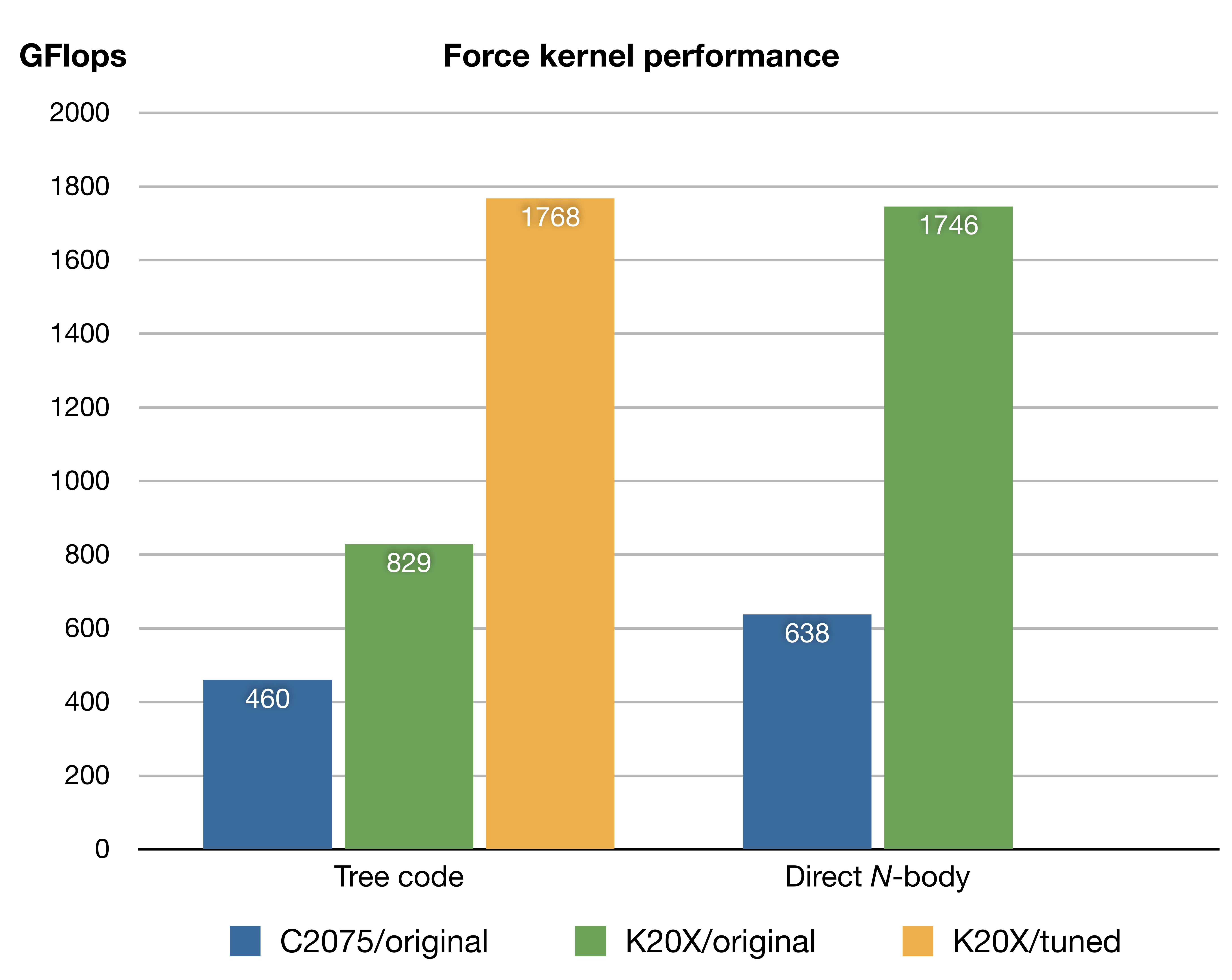}
\end{center}
\caption{Performance of the gravitational force kernel. The blue bars indicate
the performance of the Fermi kernel on the C2075. The green bars indicate the
performance of the Fermi kernel on a K20X.  The gold bar shows the performance
of the K20X tuned kernel.  With tuning, the K20X is twice as fast as the
original kernel, and is 4x faster than the C2075. Also presented here is the
performance of the direct $N$-body kernel from NVIDIA's CUDA SDk 5.5 running on
the same hardware.
\label{fig:Kepler}}
\end{figure}

The assimilation of the tree-walk and force computations into a
single GPU kernel allows us to achieve excellent computational
efficiency by 
not wasting the GPU's memory bandwidth for saving the particle interaction lists in the off-chip memory. 
Instead, the interaction lists are stored in
registers and evaluated on-the-fly during the tree-walk, therefore
delivering superb performance in excess of 1.7 Tflops on a single
K20X. The details of the tree-walk on NVIDIA Fermi architecture are
described in \cite{2012JCoPh.231.2825B}.
A naive use of the Fermi optimized kernels on Kepler GPUs delivers
relatively poor performance. In Fig.~\ref{fig:Kepler} we show that the
Fermi kernels only deliver twice the performance improvement, while
the hardware is four times faster in (peak) single precision. The main
limitation was caused by excessive use of shared memory.
The Kepler architecture introduced {\tt \_\_shfl} intrinsics for 
intra-warp\footnote{A warp is a group of threads which are executed in lock-step. 
          On current NVIDIA hardware, a warp has 32 threads.}
communication.
 By taking advantage of these capabilities, we were able to
reduce the shared memory usage by 90\% in favour of registers and recover the
missing performance.

\subsection{Multi-GPU Parallelization}\label{Sect:ParallelOptimizations}

Maintaining {\tt Bonsai}'s excellent single-GPU efficiency when scaled
to many GPUs requires both the minimization of the amount of data
traffic between different GPUs, and hiding the communication steps
behind computations. We achieve this by carefully selecting, combining
and expanding different well-known parallelization strategies. After
experimenting we eventually settled on a combination of the Local
Essential Tree (LET) and Space Filling Curve (SFC) methods.

In the original LET method~\cite{Warren:1992:ANS:147877.148090}, the physical domain is divided into
rectangular sub-domains via a recursive multi-section algorithm. Each
process uses these sub-domains to determine which part of its local
data (that is to say which LET) will be required by a remote process. 
After a process has received all the
required LET structures, they are merged into the local tree to
compute the gravitational forces. This method was used in several
award-winning papers
\cite{Hamada:2009:THN:1654059.1654123,Hamada:2010:TAN:1884643.1884644,Ishiyama:2012:PAN:2388996.2389003}.

\begin{figure}[t]
\begin{center}
\includegraphics[width=70mm]{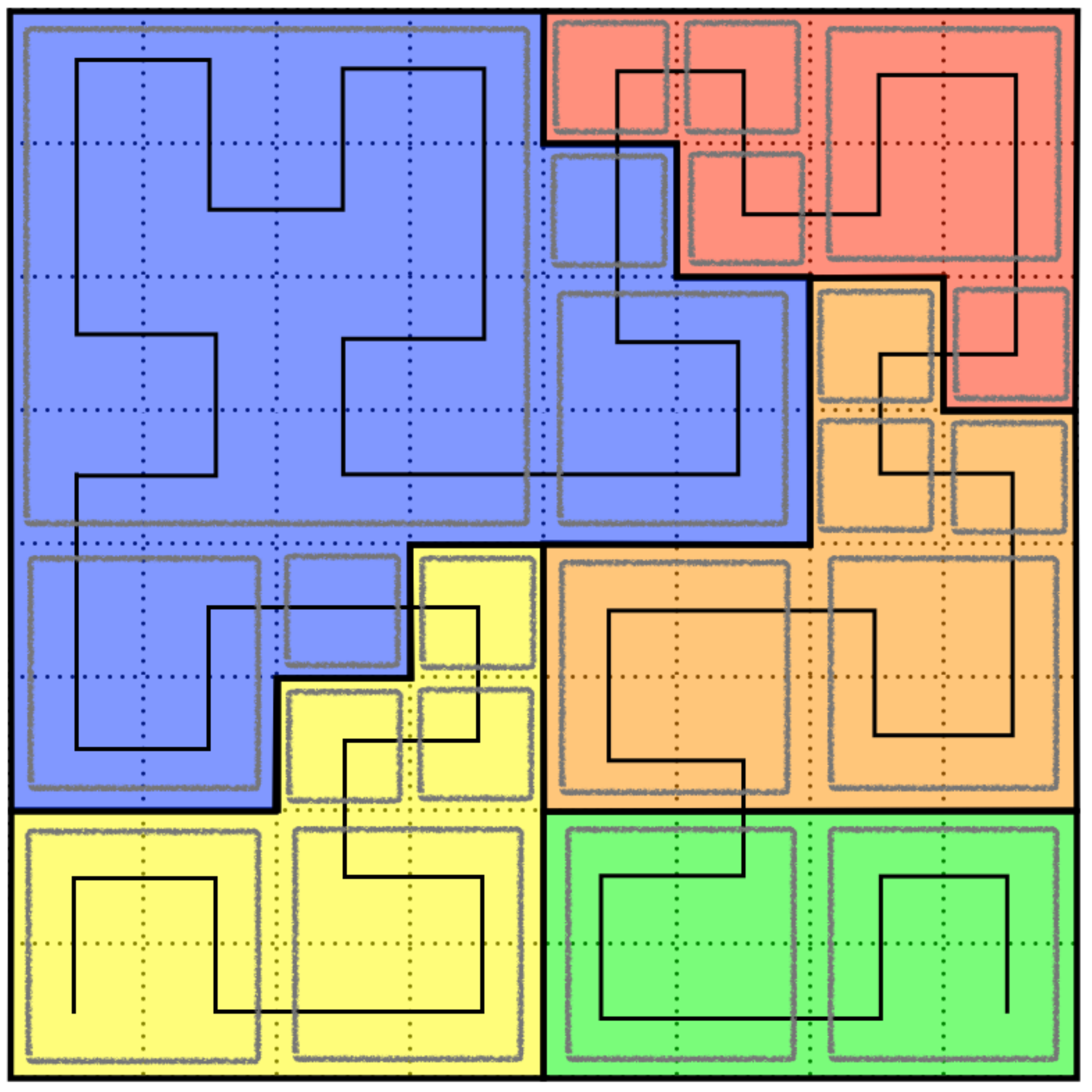}
\end{center}
\caption{Example of domain decomposition with a Peano-Hilbert space filling
curve (black solid line). The colors indicate the five separate domains.
The gray squares correspond to the tree-cells, which consist solely of
particles belonging to a single process.
\label{fig:PH}}
\end{figure}

In the SFC method, particles are ordered along an SFC which is split into equal
pieces that define the sub-domain boundaries. These boundaries will have a
fractal shape, which makes it non-trivial to build a compact LET structure.  For
example, Fig.~\ref{fig:PH} illustrates a Peano-Hilbert space filling curve
(PH-SFC)~\cite{citeulike:2861104} and the corresponding decomposition to 5
sub-domains.  To avoid these complications, SFC-based tree codes usually do not
use LET structures. Rather, they either export particles to remote processes,
compute partial forces, and import results back~\cite{2005MNRAS.364.1105S}, or
request sub-trees from the remote processes~\cite{169640,Winkel2012880,
Warren:2013:IPH:2503210.2503220}.  Such methods generate multiple communication
steps during the tree-walk.

The LET method requires the least amount of communication. For this 
reason, and despite the difficulties of dealing with the fractal boundaries, we
combine LET with the SFC domain decomposition which guarantees that sub-domain
boundaries are branches of a hypothetical global
octree. This step allows us to skip merging the imported structures
into the local-tree, but rather process them separately as soon as
they arrive, thereby hiding the communication behind computations.

\subsubsection{Domain Decomposition \& Load Balancing}

We take the following steps to compute the Peano-Hilbert (PH)
keys~\cite{citeulike:2861104} that form the basis of the tree-construction and
domain composition. Each GPU computes its local bounding box, after which the
CPUs determine the global bounding box, whose geometric properties are used to
map particle coordinates into global PH keys.  These keys are used to determine
the local domain geometry of each of the processes by means of the parallel
sampling method.

In the original sampling method~\cite{Blackston:1997:HPE:509593.509597}, each
process samples local keys with a fixed rate, $R$, which are sent to a
domain-decomposition process (DD-process) that combines these into a global
PH-SFC. This SFC is cut into $p$ equal pieces, whose beginning and ending
PH-key determines the corresponding local domain geometry of individual processes.
This method, however, proved to be inadequate for our purpose because as the
number of processes increases, and while keeping the average number of particles
per process the same, the sampling rate must be increased. The net result
is that for a sufficiently large $p$ the domain decomposition becomes a serial
bottleneck in the code.

We parallelized the original sampling method by first factoring the number of
processes, $p = p_x \times p_y$, where $p_x$ is the number of DD-processes 
and $p_y$ the number of processes handled by each DD-process.
Similarly to the original method, each process samples local keys with a fixed
sampling rate, $R_1$, which are then gathered by a DD-process. However, in contrast to
the serial method, the resulting SFC is cut into $p_x$ equal pieces, whose
boundaries are broadcast. Next, each of the processes samples keys with a
different fixed rate, $R_2$, and the sampled keys are sorted into each of the
$p_x$ sub-domains. The keys belonging to each of the sub-domains are gathered by
a corresponding DD-process, which combines them into a partial SFC and cuts
it into $p_y$ equal pieces; in total, $p_x$ DD-processes will have $p_y$ pieces
each. All these are gathered and merged into the $p = p_x \times p_y$ pieces by a
DD-process.  As in the original method, the beginning and ending PH keys of
the pieces determine the new local domain geometry for each of the processes.

Finally, each of the sampling rates includes a correction factor
needed to achieve load balance. In particular, we balance the number
of floating point operations executed by the GPU tree-walk kernel with
the restriction that a process cannot have 30\% more than the
average number of particles per GPU~\cite{2009PASJ...61.1319I}.

With the domain boundaries at hand, each GPU generates a list of
particles that are not part of its local domain, and these particles
are then exchanged between the processes. After the particle exchange is
complete, each GPU rebuilds its tree-structure and computes the
corresponding multipole moments. At the end of this step, each process
has all necessary information to proceed with the force
calculation. We would like to stress that each local tree is a
non-overlapping branch of a hypothetical global octree, because we use the PH-SFC
as a basis for the domain decomposition. This guarantees binary
consistency of the domain decomposition independent of the number of
processes, and allows us to hide LET communication behind computation.

\subsubsection{Computing the gravity}

Due to the long-range nature of Newton's universal law of gravitation
\cite{Newton:1687}, the computation of mutual forces is by definition
an all-to-all operation.  Therefore, to compute forces on local
particles, a target process requires communication with all other
processes during each iteration. We do this by forcing remote processes
to send the required particle and cell data (LETs) to the target process. 
In order to build a LET structure for a remote domain, a process 
has to know the physical boundaries of that remote domain. 

To extract these boundaries we use the local tree-structure and select the cells
that form the edges of the local particle set (gray squares in Fig.\,\ref{fig:PH}).
We then send a copy of our local tree in which all cells except these boundary
cells (and their parents) are removed. 
In this way, we can also use this tree as a LET structure.  To
gather boundary trees, we use the {\tt MPI\_Allgatherv} collective. Once this
communication step is complete we continue with the computation and exchange of
the actual LET structures. This requires two steps, both of which contain
floating point and memory bandwidth intensive operations. In order for the 
code to scale efficiently, these steps must overlap with the force
computation on the GPU.  To achieve this we use CPU multi-threading via
OpenMP, with which we split each MPI process into three thread-groups: one
thread is responsible for MPI
communication, which we refer to as the communication thread, another thread
drives the GPU, which we call the driver thread, and the rest of the threads
are busy computing, which we collectively call the compute threads.

In the first step, the compute threads check whether the boundaries of the remote
domain contain enough information to compute the forces for our local particles,
or whether a more extensive LET is required from the remote domain. At the same 
time we also do the same check whether the boundary information that we have sent
to the remote domain is indeed sufficient for that domain to compute the 
gravitational forces, or whether a more extensive LET structure has to be sent.
By carrying out the same checks for ourselves and for the remote domain we perform 
double the amount compute work but this reduces the amount of required communication
and increases the asynchronicity of the LET process, as we do not have to wait on 
a signal from the remote domain.  If the boundaries can be used as LET structures, 
no further communication will be required with the processes for which we use the
boundaries. The exceptions are our $\sim$40 nearest neighbors which 
require more data than our boundaries can provide and for these 
processes we have to execute the second step. 

In the second step, the compute threads prepare specific LET structures 
for each of the domains that require more information than 
what is contained in the boundaries. At the same time, the
communication thread is busy sending prepared LET data as it arrives
from the compute threads, as well as receiving LET data from remote
processes which it passes to the driver thread. The driver thread
merges these LET trees into a sub-tree, in which LET structures form
branches.  In other words, this reconstructs the necessary  branches of the
hypothetical global octree on the fly, but only using LET data currently
received from remote processes.  Whenever the GPU is ready with the
gravitational force kernel, either on the local data or the already received
remote data, the newly built LET tree is fed to the GPU. 

Once the gravitational forces are computed for all local particles, we advance
them forward in time using a $2^{\rm nd}$-order
leap-frog integration scheme \cite{1995ApJ...443L..93H}. This marks
the end of the $N$-body integration step.

Since the above method is based on the tree-structures, the amount of
communication per process during the boundary exchange is virtually independent
of the number of particles per GPU, and depends only on the number of processes
and the simulated physical model. 
Furthermore, the gravity step as a whole becomes more efficient with more
particles per GPU because the time window to hide the communication becomes
larger.  The communication time itself increases only slightly because the
number of particles at the domain surface, which are commonly used for the communication,
increases at a lower rate than the total number of particles inside the domain
volume.

\section{Application and Early Scientific Results} 
\label{Sect:4}

\begin{figure*}
\begin{center}
\includegraphics[width=0.8\textwidth,angle=0]{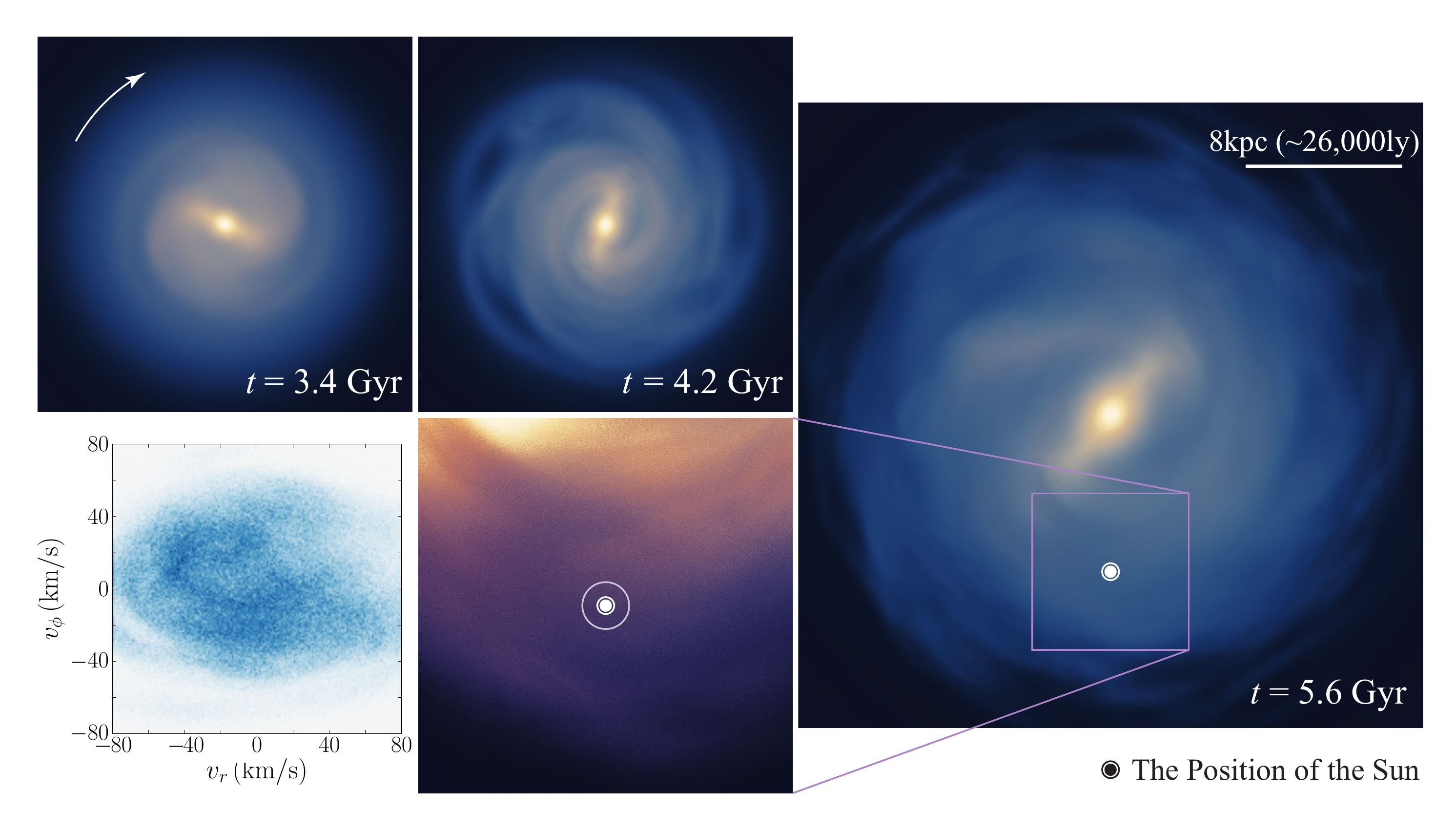}
\end{center}
\caption{Face-on surface density of the Galactic disk and the bulge from the 51
billion particle Milky Way simulation.  The two top panels show the density at
3.4 and 4.2 billion years, and the arrow in the left panel shows the rotation
direction of the Galactic disk. The large rightmost panel shows density plot at 5.6 billion years.
The bottom right panel is a zoom-in of the spiral region marked in the rightmost
panel, where the Sun in the Milky Way is thought to be located (8 kpc from the
Galactic Center).  The bottom left panel is the radial and azimuthal velocity
distribution of the stars within a 500 pc radius centered around the position
of the Sun (circle marker in the bottom middle panel), in which the rotation velocity
of the disk is subtracted from the azimuthal velocity.  
\label{fig:MW}}
\end{figure*}

The Milky Way model that we use contains a central bulge and stellar disk 
which are embedded inside a dark matter halo.
In this model~\cite{2008ApJ...679.1239W} the halo has a mass
of $6.0\times10^{11}M_{\odot}$ and uses an Navarro-Frenk-White (NFW) density 
distribution~\cite{1996ApJ...462..563N}. The stellar disk has a mass 
of $5.0\times 10^{10}M_{\odot}$ and follows an exponential distribution
while the bulge stars are based on a Hernquist density profile \cite{1990ApJ...356..359H} 
with a total mass of $4.6\times 10^9M_{\odot}$. 
We realized this model with a total of 51,199,967,232 (51 billion) particles,
divided over the bulge, disk, and halo with 994,689,024 (1 billion),
2,945,105,920 (3 billion), and 47,260,172,288 (47 billion) particles respectively. 
We adopt equal masses for each of the particles for all three components in order 
to avoid numerical heating caused by unequal mass.
This results in
a mass resolution of $\sim 10 M_{\odot}$, and a spatial resolution of 1 
parsec\footnote{1 parsec is 3.26 light year.} (pc).

This simulation has a particle resolution in the bulge and disk
200 times greather than the previous highest resolution
simulation of the Milky Way~\cite{2014arXiv1404.0304D}.
The halo is resolved with 500 and 5000 times more particles compared 
to~\cite{2014arXiv1404.0304D} and~\cite{2009ApJ...703.2068D} respectively.
To generate the initial conditions (IC) for the Milky Way model, we use 
GalacticICS~\cite{2005ApJ...631..838W} which creates a smooth exponential disk,
and a spherically symmetric bulge and halo. Given the sheer size of the initial
model, it is crucial to modify this IC generator for distributed and
multi-threaded execution. 
Furthermore, to avoid heavy start-up I/O related to reading large 
IC files, we decided to generate all our Milky Way models on the fly.

The large number of particles enables us to adopt a smaller softening
length, which results in significantly higher spatial resolution compared to
previous simulations; in contrast to mass resolution, spatial resolution scales as
${\cal O}(N^{1/3})$. We adopt a softening length of 1 pc, 
which is an order of magnitude smaller than in previous simulations~\cite{2009ApJ...703.2068D, 2014arXiv1404.0304D}. 
Furthermore most of the previous simulations adopted an opening angle for the tree-code 
of $\theta=0.7$, which is acceptable for
spheroidal systems, but to accurately model irregular small-scale structures, such as spiral arms,
a significantly smaller opening angle is required. Therefore we used $\theta=0.4$
even though this increases the calculation
costs proportionally to ${\cal O}(\theta^{-3})$ \cite{1991PASJ...43..621M}.  
We simulated this Milky Way model for 6 billion years, at which point spiral
arms and a bar structure have fully formed.

In the top panels of Fig.\,\ref{fig:MW} we present the time evolution 
of the face-on surface density of the disk and bulge. 
The galaxy did not form any prominent structure up to half-way through
the simulation (${\sim}3$  billion year). Shortly afterward, however, a bar structure
formed which then induced the formation of spiral arms. At $t\sim 5$ billion years, a barred spiral galaxy similar to
the Milky Way has formed. The time at which the bar and spiral structure 
forms appears to increase with the number of particles, and therefore a longer 
simulation time may be required for larger models.
The bottom right panel of Fig.\,\ref{fig:MW} is a zoom-in of the area around where 
the Solar System is located, 8 kpc ($\sim$26,000 light-year) from the  Galactic Center.
The high resolution of the simulations allows us to resolve the fine structure of the 
spiral arms and the results can be directly compared to observations.

In the bottom left panel of Fig.\,\ref{fig:MW}, we present the velocity structure near the Sun,
which is the region marked by the circle in the bottom middle panel. Several streams and spots 
of high density regions can be observed. These are also known as `moving groups', and many similar structures 
have been found in recent observations of the stars in the Solar Neighborhood\cite{2012MNRAS.426L...1A}.
These moving group structures are considered to be formed by resonances 
originating from the bar \cite{2000AJ....119..800D}; however, also
the spiral arms are capable of forming such velocity structures
\cite{2011MNRAS.418.1423A}. Comparing our simulation data with 
observational data allows us to study the dynamics of
the bar and the spiral structures in the Milky Way.
With this simulation we are able, for the first time, 
to obtain enough sample particles (68,000 within 500 pc) from a self-consistent 
simulation to make a direct comparison with observations that 
contain a comparable number of observed stars (57,000 within 200--300 pc) \cite{2012MNRAS.426L...1A}.

At the time of writing, observations have only provided kinematic data for the Milk Way disk stars
that are located in the Solar Neighborhood \cite{2012MNRAS.426L...1A}.
However, the future Gaia catalogue will extend this to stars beyond the Solar
Neighborhood, and comparing these observations against our simulations will
provide us with further details of the structure of the Milky Way Galaxy 
such as the size and the pattern speed of the bar, the structure of 
the spiral arms, and the resonance effects in the velocity distribution.

In addition, a more detailed analysis of the currently simulated galaxy 
will provide us with numerous pieces of information that can point us
to the origin of the observed structures in the Milky Way and 
how it has evolved during the 10 billion years that it took to form 
the current-day Milky Way.

\section{System and environment where \\performance was measured} 
\label{Sect:5}

The {\tt Bonsai} tree-code we use to perform our Milky Way simulations is optimized to run
on the NVIDIA Fermi and Kepler architectures (see \S\,\ref{Sect:3}). The highest performance
is obtained on the K20X, because of the hardware
improvements introduced with the Kepler generation GPUs. The code
operates on distributed memory systems with one or more GPUs per node. 

For the performance measurements and production runs in this work we used two
machines. The first is the Swiss Cray XC30 machine, Piz Daint, which has 
a total of 5272 nodes. The second is the Cray XK7 supercomputer, Titan, at Oak
Ridge National
Laboratory in the USA which has a total of 18688 nodes (see Table~\ref{tab:Hardware} for details).
Although both these machines are designed by Cray, they have different
interconnect and CPUs.  
Piz Daint is the newer of the two supercomputers, and is equipped with Intel
CPUs as well as a more advanced network generation, Aries, where the nodes are connected in
a dragonfly topology.  Titan, on the other hand, contains AMD CPUs and the
nodes are connected using the Gemini network architecture which uses a 3D torus
topology. The faster CPUs and much improved interconnect of Piz Daint benefit
the parallel performance of {\tt Bonsai} during the communication phase as we
will see in \S\,\ref{Sect:6}.

{\tt Bonsai} runs on any computer as long as it contains an NVIDIA GPU with CUDA support.
This includes laptops, desktops and clusters. The number of particles that can be used 
per GPU depends on the amount of memory available to the GPU. Both the Fermi and Kepler 
architectures are supported. For single GPU simulations the only CPU communication required
involves kernel launches and the retrieval of results from the GPU. When using multiple 
GPUs the speed of the CPU, PCI-E bus and network become crucial to achieve high 
performance as we have to communicate between all the MPI processes. The CPU cores are 
used for writing result data to disk, network communication, preparing LET data 
to be sent over the network, receiving LET data that is sent to us by a remote 
process, and for feeding the GPU. The GPU handles the data processing, 
force computations, particle integration and creates the tree-structures for 
the local data.

\begin{table}[!t]
\caption{
  Hardware used for our parallel simulations.
  On both systems we used CUDA 5.5, GCC 4.8.2
  and the Cray MPICH 6.2 library.
}
\label{tab:Hardware}
\centering
\vspace{0.75em}
\begin{tabular}{|lcc|}
\hline
Setup           & Piz Daint & Titan \\
\hline
GPU model       & K20X   & K20X \\
GPU/node        &    1    &    1 \\
Total GPUs       & 5272    & 18688 \\
GPUs used       & 5200    & 18600 \\
GPU RAM (ECC enabled)        & 5.4 GB & 5.4 GB\\
\hline
CPU model       & Xeon E5-2670 & Opteron 6274  \\
CPU/node        & 1     & 1 \\
Total CPUs       & 5272   & 18688 \\
CPUs used       & 5200   & 18600 \\
CPU cores used  & 41600 & 297600 \\
Node RAM        &  32GB  &  32 GB\\
\hline 
Network         & Cray Aries/dragonfly & Cray Gemini/3D Torus \\
\hline
\end{tabular}
\end{table}

\section{Performance results}\label{Sect:Performance}
\label{Sect:6}

We present the performance results by calculating floating point
operations for \emph{only} the force calculations. We ignore contributions
from tree construction, computation of multipole moments, multipole
acceptance criteria during the tree-walk, all LET-related operations
(which are performed on the CPU), time integration and diagnostics. The timing of these operations 
however, is taken into account when we calculate the application performance.

\subsection{Operation counts}

The accelerations, $\bm a_i$, and potential, $\phi_i$, of particle $i$, 
which we collectively call the force, are computed by
\newcommand{\trQ }{{\rm tr}({\bf Q}_j)}
\newcommand{\Qr  }{{\bf Q}_j {\bm r_{ij}}}
\newcommand{\rQr }{{\bm r_{ij}^T}{\bf Q}_j {\bm r_{ij}}}
\newcommand{\rabs}{|\bm r_{ij}|}
\newcommand{\rij }{\bm r_{ij}}
\begin{eqnarray}
\phi_i &=& \sum_j \left[ 
    -\frac{m_j}{\rabs}
    +\frac12 \frac{\trQ}{\rabs^3} 
    -\frac32 \frac{\rQr}{\rabs^5}
    \right],  \\
\bm a_i &=& \sum_j \left[
      \frac{m_j \rij}{\rabs^3} 
      - \frac32 \frac{\trQ \rij}{\rabs^5}
      \right. \nonumber \\ && \left. \quad\quad
      - \frac{3\Qr}{\rabs^5}
      + \frac{15}2 \frac{(\rQr)\rij}{\rabs^7}
    \right], 
\end{eqnarray}
where $\bm r_{ij} = \bm r_j - \bm r_i$.  Here, $m_j$, $\bm r_j$, ${\bf
  Q}_j$ are mass, position and quadrupole moments (in a $3\times3$
symmetric matrix) of particle $j$ respectively. Computation of one
particle-particle interaction (p-p) without the quadrupole moment term
consists of 4 subtraction (sub), 3 multiplication (mul), 6
fused-multiply-add (fma), and 1 reciprocal-square-root (rsqrt)
instructions. In this article, we count 4 floating-point operations
for the reciprocal-square-root, which results in a total of {\bf 23}
operations for each p-p. A particle-cell interaction (p-c) with quadrupole
corrections consists of 4 sub, 6 add, 17 mul, 17 fma and 1 rsqrt,
which results in {\bf 65} operations for p-c interaction\footnote{ All
  operation counts were verified with the disassembling command {\tt
    cuobjdump -sass} in the CUDA toolkit.}.  The total number of flops
is obtained by multiplying these numbers by the total number of p-p
and p-c (as recorded during execution), and divided by the execution
time.

Note that
\cite{Warren:1992:ANS:147877.148090,Warren:1998:AAC:509058.509130,
  Kawai:1999:ANB:331532.331598, Hamada:2009:THN:1654059.1654123,
  Hamada:2010:TAN:1884643.1884644} used {\bf 38} for the operation
count of a p-p interaction.  Although it is convenient to use the same
operation count for comparing one record with the other, this can
over count the operations for hardware with fast {\tt rsqrt} support.  
Previous GBP winner \cite{Ishiyama:2012:PAN:2388996.2389003} counted {\bf
  51}, within which about half was spent in the calculation of a
cut-off polynomial.

\subsection{Scalability}

To evaluate the scalability and parallel performance of {\tt Bonsai} we conduct
both weak and strong scaling studies. In both studies we use the Milky Way
Galaxy model as initial condition (see \S\,\ref{Sect:4}), the model was
integrated for 64 time-steps. The measurements to assess code scalability were
taken from time-steps 32 until 64, which allowed the simulation to relax into a
load-balanced state. This is a good representation of long-term integrations,
in which significant changes to the physical configuration take place on a
time-scale of hundreds of iterations.

In the weak scaling test we used on average 13 million particles per GPU, which
is comparable to the number of particles we used in the production runs. It is
possible to do runs with up to 20 million particles per K20X, and thereby achieve higher
application performance, as more time is spent on the GPU, but that is not
comparable to the production run configurations that we use.  For all the
performance measurements and production simulations we used an opening angle of
$\theta=0.4$, which is satisfactory to properly model disk galaxies, such as
the Milky Way (see \S\,\ref{Sect:4}).

\begin{figure*}
\begin{center}
\includegraphics[width=0.6\textwidth,angle=-90]{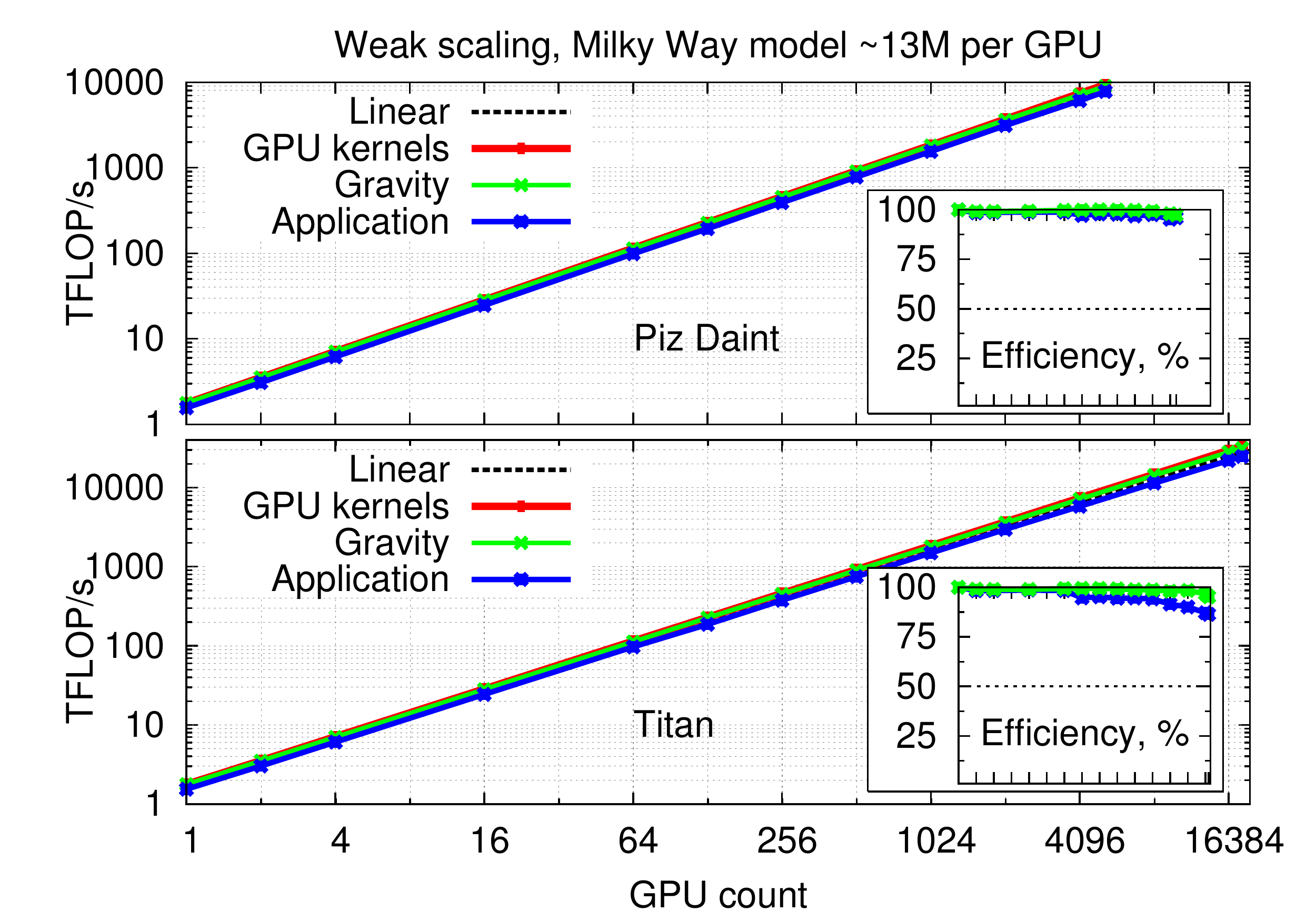}
\end{center}
\caption{Weak scaling performance for Piz Daint and Titan. 
  For both panels we used simulations of our Milky Way model 
  with on average 13 million particles per GPU 
  and $\theta=0.4$ as the opening angle.
  The x-axis of the panels indicates the number of GPUs used and 
  the y-axis the performance in Tflops. 
  The red solid lines show the performance of the tree-walk
  kernel running on the GPU. The green solid lines show the
  performance of the gravity step, including the time to communicate
  LET structures. The blue solid lines show the performance of the
  full application.
  The black dashed lines, which are mostly hidden behind the blue 
  lines, indicate linear scaling
  The lower right insets shows parallel efficiency of the code when compared
  to a single node. In these insets the green line indicates the efficiency
  of the gravity step, including communication. The blue line indicates
  parallel application efficiency with respect to a single GPU.  
  The top panel shows the performance of the simulations on 
  Piz Daint for 1 to 5200 GPUs. 
  The bottom panel shows the performance of the simulations on 
  Titan for 1 to 18600 GPUs. 
\label{fig:ScalingWeak}}
\end{figure*}

In Fig.\,\ref{fig:ScalingWeak} we show the weak scaling and parallel efficiency
for simulations on both Titan and Piz Daint.  On Piz Daint we used between 1
and 5200 GPUs and on Titan we used between 1 and 18600 GPUs. The parallel
efficiency of the simulations on Piz Daint never drops below 95\%, indicating
that most of the required communication is hidden behind the GPU work. On Titan
the parallel
efficiency for the simulations up to 8192 GPUs is around 90\%. This is slightly
lower than on Piz Daint and is caused by a combination of the slightly higher
communication times on the older network and slightly longer LET generation
times because of the slower AMD CPUs. The breakdown of these timings can be
found in Tab.~\ref{Tab:timeBreakdown}. For the largest run on Titan, with 18600
nodes and a total of 242 billion particles, we reach a GPU single precision
performance of 33.49 Pflops and a sustained application performance of 24.77
Pflops. This results in a parallel efficiency of 86\% when compared to a single
GPU. The drop in efficiency when compared to 8192 nodes is due to both longer
network communication times and increased imbalance. Finally, on 18600 nodes
with 242 billion particles in total, the average wall-clock time of
a full time-step is 4.8 seconds.
\JB{TODO, import new 8k strong it16 data }

We also demonstrate the strong scaling behavior of {\tt Bonsai} in
Tab.~\ref{Tab:timeBreakdown}. In particular, on Piz Daint we simulated a 26.6
billion particle Milky Way model, while on Titan we simulated a model two 
times larger. In both cases, we find
satisfactory strong scaling, with a parallel efficiency of 95\% on Piz Daint 
when scaled from 2048 to 4096 GPUs and 87\% on Titan when scaled from
4096 to 8192 GPUs.

\begin{table*}
\begin{center}
\caption{$\mathrm{Time}$ $\mathrm{breakdown}$ $\mathrm{for}$ $\mathrm{Titan}$ $\mathrm{and}$ $\mathrm{Piz}$ $\mathrm{Daint}$ $\mathrm{using}$ $\mathrm{different}$ $\mathrm{numbers}$ $\mathrm{of}$ 
$\mathrm{GPUs}$ $\mathrm{and}$ $\mathrm{particles.}$ $\mathrm{The}$ $\mathrm{number}$ $\mathrm{of}$ $\mathrm{particles}$ $\mathrm{per}$ $\mathrm{GPU}$ $\mathrm{used}$ $\mathrm{in}$ $\mathrm{the}$ $\mathrm{simulations}$ $\mathrm{is}$ 
$\mathrm{indicated}$ $\mathrm{by}$ $\mathrm{the}$ $\mathrm{first}$ $\mathrm{row}$ $\mathrm{of}$ $\mathrm{the}$ $\mathrm{table.}$ $\mathrm{For}$ $\mathrm{each}$ $\mathrm{simulation}$ $\mathrm{we}$ 
$\mathrm{use}$ $\mathrm{our}$ $\mathrm{Milky}$ $\mathrm{Way}$ $\mathrm{model}$ $\mathrm{as}$ $\mathrm{input}$ $\mathrm{data}$ $\mathrm{and}$ $\theta=0.4.$
$\mathrm{The}$ $\mathrm{data}$ $\mathrm{presented}$ $\mathrm{shows}$ $\mathrm{the}$ $\mathrm{major}$ $\mathrm{parts}$ $\mathrm{of}$ $\mathrm{the}$ $\mathrm{algorithm}$ $\mathrm{and}$
$\mathrm{by}$ $\mathrm{which}$ $\mathrm{computation}$ $\mathrm{unit}$ $\mathrm{they}$ $\mathrm{are}$ $\mathrm{primarily}$ $\mathrm{executed.}$
$\mathrm{For}$ $\mathrm{the}$ $\mathrm{first}$ $\mathrm{set}$ $\mathrm{of}$ $\mathrm{rows}$ $\mathrm{the}$ $\mathrm{first}$ $\mathrm{column}$ $\mathrm{indicates}$ $\mathrm{the}$ $\mathrm{number}$ $\mathrm{of}$ 
$\mathrm{particles}$ $\mathrm{and}$ $\mathrm{the}$ $\mathrm{part}$ $\mathrm{of}$ $\mathrm{the}$ $\mathrm{algorithm}$ $\mathrm{that}$ $\mathrm{is}$ $\mathrm{involved.}$ $\mathrm{The}$ $\mathrm{second}$ $\mathrm{and}$
$\mathrm{third}$ $\mathrm{column}$ $\mathrm{show}$ $\mathrm{the}$ $\mathrm{computation}$ $\mathrm{unit}$ $\mathrm{which}$ $\mathrm{is}$ $\mathrm{involved}$ $\mathrm{in}$ $\mathrm{this}$ $\mathrm{part}$ 
$\mathrm{of}$ $\mathrm{the}$ $\mathrm{algorithm.}$ $\mathrm{The}$ $\mathrm{following}$ $\mathrm{columns}$ $\mathrm{show}$ $\mathrm{the}$ $\mathrm{execution}$ $\mathrm{time,}$ $\mathrm{where}$ $\mathrm{the}$ 
$\mathrm{first}$ $\mathrm{column}$ $\mathrm{shows}$ $\mathrm{the}$ $\mathrm{single}$ $\mathrm{GPU}$ $\mathrm{timings,}$ $\mathrm{the}$ $\mathrm{following}$ $\mathrm{five}$ $\mathrm{the}$ $\mathrm{timings}$ $\mathrm{on}$ 
$\mathrm{Titan}$ $\mathrm{and}$ $\mathrm{the}$ $\mathrm{final}$ $\mathrm{four}$ $\mathrm{the}$ $\mathrm{timings}$ $\mathrm{on}$ $\mathrm{Piz}$ $\mathrm{Daint.}$ $\mathrm{Note}$ $\mathrm{that}$ $\mathrm{the}$ $\mathrm{fifth}$ $\mathrm{and}$ $\mathrm{fourth}$ $\mathrm{columns}$
$\mathrm{of}$ $\mathrm{the}$ $\mathrm{Titan}$ $\mathrm{and}$ $\mathrm{Piz}$ $\mathrm{Daint}$ $\mathrm{data}$ $\mathrm{respectively}$ $\mathrm{shows}$ $\mathrm{runs}$ $\mathrm{with}$ $\mathrm{fewer}$ $\mathrm{particles}$ $\mathrm{to}$ $\mathrm{present}$ $\mathrm{the}$ 
$\mathrm{strong}$ $\mathrm{scaling}$ $\mathrm{data.}$ 
$\mathrm{The}$ `$\mathrm{Non}$-$\mathrm{hidden}$ $\mathrm{LET}$ $\mathrm{comm}$' $\mathrm{row}$ $\mathrm{indicates}$ $\mathrm{the}$ $\mathrm{part}$ $\mathrm{of}$ $\mathrm{the}$ $\mathrm{communication}$ $\mathrm{time}$ 
$\mathrm{during}$ $\mathrm{the}$ $\mathrm{gravity}$ $\mathrm{step}$ $\mathrm{that}$ $\mathrm{was}$ $\mathrm{not}$ $\mathrm{hidden}$ $\mathrm{behind}$ $\mathrm{the}$ $\mathrm{GPU}$ $\mathrm{work.}$
$\mathrm{The}$ `$\mathrm{Other}$' $\mathrm{data}$ $\mathrm{includes}$ $\mathrm{the}$ $\mathrm{time}$ $\mathrm{to}$ $\mathrm{allocate}$ $\mathrm{memory,}$ $\mathrm{gather}$ $\mathrm{and}$ $\mathrm{write}$ 
$\mathrm{statistics,}$ $\mathrm{integrate}$ $\mathrm{particles}$ $\mathrm{and}$ $\mathrm{any}$ $\mathrm{waiting}$ $\mathrm{times}$ $\mathrm{induced}$ $\mathrm{by}$ $\mathrm{load}$-$\mathrm{imbalance.}$ $\mathrm{}$ 
$\mathrm{The}$ $\mathrm{second}$ $\mathrm{set}$ $\mathrm{of}$ $\mathrm{two}$ $\mathrm{rows}$ $\mathrm{presents}$ 
$\mathrm{the}$ $\mathrm{average}$ $\mathrm{number}$ $\mathrm{of}$ $\mathrm{particle}$-$\mathrm{particle}$ $\mathrm{and}$ $\mathrm{particle}$-$\mathrm{cell}$
$\mathrm{interactions}$ $\mathrm{as}$ $\mathrm{recorded}$ $\mathrm{during}$ $\mathrm{the}$ $\mathrm{simulations.}$
$\mathrm{The}$ $\mathrm{bottom}$ $\mathrm{two}$ $\mathrm{rows}$ $\mathrm{give}$ $\mathrm{the}$ $\mathrm{resulting}$ $\mathrm{performance}$ $\mathrm{in}$ $\mathrm{Tflops}$ 
$\mathrm{for}$ $\mathrm{the}$ $\mathrm{GPU}$ $\mathrm{kernels}$ $\mathrm{(excluding}$ $\mathrm{communication)}$ $\mathrm{and}$ $\mathrm{the}$ $\mathrm{application}$ $\mathrm{as}$ $\mathrm{a}$ $\mathrm{whole.}$ 
$\mathrm{For}$ $\mathrm{Titan}$ $\mathrm{we}$ $\mathrm{present}$ $\mathrm{the}$ $\mathrm{weak}$ $\mathrm{scaling}$ $\mathrm{results}$ $\mathrm{for}$ $\mathrm{1024,}$ $\mathrm{2048,}$ $\mathrm{4096}$ $\mathrm{and}$ $\mathrm{18600}$ $\mathrm{GPUs}$ $\mathrm{and}$
$\mathrm{the}$ $\mathrm{strong}$ $\mathrm{scaling}$ $\mathrm{data}$ $\mathrm{for}$ $\mathrm{4096}$ $\mathrm{and}$ $\mathrm{8192}$ $\mathrm{GPUs.}$ $\mathrm{}$
$\mathrm{For}$ $\mathrm{Piz}$ $\mathrm{Daint}$ $\mathrm{we}$ $\mathrm{present}$ $\mathrm{the}$ $\mathrm{weak}$ $\mathrm{scaling}$ $\mathrm{results}$ $\mathrm{for}$ $\mathrm{1024,}$ $\mathrm{2048}$ $\mathrm{and}$ $\mathrm{4096}$ $\mathrm{GPUs}$ $\mathrm{and}$
$\mathrm{the}$ $\mathrm{strong}$ $\mathrm{scaling}$ $\mathrm{data}$ $\mathrm{for}$ $\mathrm{2048}$ $\mathrm{and}$ $\mathrm{4096}$ $\mathrm{GPUs.}$ 
}
\label{Tab:timeBreakdown}
\vspace{0.75em}
\begin{tabular}{|l|ccc|ccccc|ccccc|}
\hline
Operation           & \multicolumn{2}{|c}{Compute unit} & time [s]
                         & \multicolumn{5}{|c}{wall-clock time Titan [s]} & \multicolumn{4}{|c}{wall-clock Time Piz Daint [s]} & \\
                    & CPU & GPU        & 1 GPU  & 1024  & 2048 & 4096  & 18600 & 8192  & 1024 & 2048  & 4096 & 4096 & \\
\hline
Nparticles/GPU [Million]   & --- & --- &  13    & 13    & 13   & 13    & 13    & 6.5   & 13   & 13    & 13   & 6.5  & \\
Sorting SFC                & --- & X   &  0.1   & 0.1   & 0.1   & 0.1   & 0.13  & 0.06  & 0.1  & 0.10  & 0.1  & 0.05 & \\
Domain Update              &  X  & --- &  ---   & 0.2   & 0.2   & 0.2   & 0.3  & 0.15  & 0.1  & 0.10  & 0.1  & 0.07 & \\
Tree-construction          & --- & X   &  0.11  & 0.1   & 0.1   & 0.1   & 0.1   & 0.05  & 0.1  & 0.10  & 0.1  & 0.05 & \\
Tree-properties            & --- & X   &  0.03  & 0.03  & 0.03  & 0.036 & 0.03  & 0.016 & 0.03 & 0.03  & 0.03 & 0.016  & \\
Compute gravity Local-tree & --- & X   &  2.45  & 1.45  & 1.45  & 1.45  & 1.45  & 0.68  & 1.45 & 1.45  & 1.45 & 0.68 & \\
Compute gravity LETs       & --- & X   &  ---   & 1.78  & 1.89  & 2.0   & 2.09  & 1.13  & 1.79 & 1.89  & 2.02  & 1.01 & \\
Non-hidden LET comm        &  X  & X   &  ---   & 0.09  & 0.1   & 0.14  & 0.22  & 0.25  & 0.09 & 0.06  & 0.07 & 0.07 & \\ 
Unbalance + Other          &  X  & X   &  0.1   & 0.27  & 0.28  & 0.40   & 0.45  & 0.31  & 0.22 & 0.21  & 0.28 & 0.15 & \\ 
\hline
Total                      &     &     &  2.79  & 4.02  & 4.15  & 4.41 & 4.77  & 2.65  & 3.84 & 3.94  & 4.15 & 2.1 & \\
\hline\hline
Interaction type  &  &  & \multicolumn{10}{c}{interaction count per particle} & \\
\hline
Particle-Particle &  &                 & 1745   & 1715  & 1716 & 1718   & 1716 & 1716   & 1716 & 1716 & 1718 & 1714 & \\
Particle-Cell     &  &                 & 4529   & 6287  & 6527 & 6765   & 6920 & 7096  & 6290 & 6515 & 6810  & 6616 & \\
\hline\hline
Performance [Tflops]  &  & \multicolumn{2}{r}{K20X}  & \multicolumn{5}{|c}{Titan} & \multicolumn{5}{|c|}{Piz Daint}\\
\hline
GPU         &  &                      &  1.77 & 1844.6& 3693.7 & 7396.8  & {\bf 33490} & 14714 & 1844.7 & 3693.9 & 7396.9 & 7383.5 & \\
Application &  &                      &  1.55 & 1484.6& 2971.8 & 5784.9  & {\bf 24773} & 10051 & 1551.9 & 3129.9 & 6180.7 & 5947.9 & \\
\hline
\end{tabular}
\end{center}
\end{table*}

\subsection{Time-to-solution}

Our production simulation of the Milky Way Galaxy for 6 billion years with a 51
billion particle model was carried out on 4096 GPU nodes of Piz Daint.  With
this simulation we were able, for the first time, to increase the spatial
accuracy by an order of magnitude compared to previous state-of-the-art
simulations. We achieved this by using over 1000 times more particles than
before, but we also tripled the time evolution of the simulation up to 6 billion years.

After the bar structure and spiral arms are fully formed we measure the 
simulation time for 1000 iterations; the time-step duration
is determined as the average of those 1000 iterations. We measure after
the formation of the bar and spiral arms because the additional small-scale structure results in 
regions with higher density than the mean density of the model, and it is not uncommon for
gravitational tree-codes to increase the interaction count in such regions,
which negatively impacts the runtime. Indeed, the average calculation time for an
iteration at $T=3.8$ billion year is 4.6 seconds, which is about 10\% larger than at the beginning of the
simulation. In addition, there was a few percent I/O-related overhead 
related to storing intermediate simulation snapshots (for the dual purpose of
restarting and detailed analysis), as well as rudimentary on-the-fly data
analysis.

Using this data we can estimate the wall-clock
time needed to simulate the Milky Way Galaxy with a 242 billion particle model
on 18600 GPUs. With a softening of 1\,parsec, the
minimal time step required for an accurate simulation is $75,000$\,year (this
corresponds to the time that two particles pass each other within a softening
length). This softening length was also used in our earlier calculations
\cite{2013MNRAS.431..767B}, and is considerably smaller than typically used in
other Galaxy simulations
\cite{2009ApJ...697..293D,2011ApJ...730..109F,2014arXiv1404.0304D}.
If we wish to simulate the evolution of the Milky Way
Galaxy for 8\,Gyr, with a time step of 0.075\,Myr we require about 106,667
time steps.
With an expected maximum of about 5.5 seconds per time
step for a Milky Model with a central bar and fully formed spiral arms,
it will take at most a week of computing time.
For a more modest model with 106 billion
particles using 8192 nodes the maximum expected calculation time per step will
be 5.1 seconds and a full simulation, using the above settings, would take
just over six days.

\subsection{Peak performance}

We achieved a sustained application single precision performance of 24.77
Pflops on a 242-billion-particle Milky Way model with 18600 GPUs. When busy
with the force computation, the GPUs were processing at an aggregated rate of
33.49 Pflops. This translates to 1.8 Tflops per GPU and 1.33 Tflops
for the overall application performance per node. The theoretical peak single
precision performance of 18600 K20X GPUs is 73.2 Pflops.  During the force
calculation the GPUs operate at 46\% of this number, while the overall
application reached 34\% of this peak performance.

\section{Implications for future\\ systems and applications}
\label{Sect:7}

The combination of high-performance accelerators and multi-core processors in a
single compute node is one of the important trends in high performance
computing. At the time of writing, nearly 90\% of the peak performance is due
to accelerators in the two fastest supercomputers in the world. It is not
unreasonable to expect that future large-scale scientific simulations will be
tailored to run on such hybrid architectures.

The simulations of the Milky Way Galaxy on the Titan and Piz Daint supercomputers,
which we present in this paper, clearly demonstrate that hybrid architectures
are well suited for scientific calculations. However, to achieve this goal we
had to completely redesign our $N$-body tree-code to fully take advantage of the
high-performance network and parallelism offered by GPUs and CPUs.

Much of our efforts in writing {\tt Bonsai} went into multi-GPU
parallelization. Instead of letting the CPUs sit idle, we
utilized its processing power for the work required to minimize the amount of
communication, which in turn permitted us to hide most of the communication
behind GPU computations. This can be favorable compared to IBM's BG/Q system
with the dedicated communication core: the CPU was in the role
of the dedicated communication hardware, the GPU was in the role of the
processor doing actual calculations, and with several clever techniques we were
able to ensure that both were carrying out their work as concurrently as
possible.

With future systems moving toward exascale computing,
communication is expected to continue to limit performance. Vendors have
already begun to adopt Unified Memory Architectures and Direct Memory Access
inside the compute nodes to reduce the communication pressure during these
data transfers. This, however, does little to reduce limitations due to
interconnect latencies and bandwidth.  With intra-node and on-chip bandwidths
being an order of magnitude larger than that of interconnects, careful
attention should be paid to minimize communication.  In {\tt Bonsai} we
accomplish this by reducing the number of MPI calls by switching to a push
instead of pull method, and using the boundary data for creating LET structures
and for computing forces. With the increase of the GPU on-board memory (e.g.
12GB on a K40 card), we will be able to store more data on the device, which
decreases the relative amount of data that needs to be communicated between the
GPUs, thereby having more possibilities to hide the communication time behind
computations.  Without major improvement of the network latencies and bandwidth
in future hardware this will likely be the only option to maintain scalability.

With future techniques, such as the recently announced NVIDIA's NVLINK
technology, it will be possible to have much faster communication between GPUs
in the same physical node.  For {\tt Bonsai} this could mean that by careful
placement of the MPI ranks we can communicate with our direct neighbors in
particle space using this high speed connection. This, again, will reduce the
pressure on the interconnect, because less data will have to be communicated
between nodes.

The large number of cores on modern day CPUs could enable us to incorporate more
physics in future simulations.
Currently we use as many the CPU cores as possible for the generation of 
LET structures, but in the future we can reserve some of the cores for other tasks. 
The galaxy simulations could then be enriched with, for example, stellar 
evolution and massive black holes with their stellar cusps. 
The gravitational interactions around the black holes require the accuracy of a 
direct $N$-body code which, together with the stellar evolution code,
would be running on the CPU while the tree-code would be running on the GPU.
Such a combination of physics could be realized
via the decomposition of physical elements, as is realized in the Astronomical
Multi-purpose Software Environment (AMUSE)
\cite{2009NewA...14..369P,2013CoPhC.183..456P}. In earlier successful attempts
the AMUSE environment was used together with heterogeneous hardware to
distribute multi-physics simulations across a national network of specialized
computers \cite{DrostMMBPZKDS12}.

Our demonstration of the efficiency of running on a massively parallel
heterogeneous architecture indicates that hybrid hardware can give excellent
performance. 
Our achieved performance, the good scaling and the short
time-to-solution have been realized by a fundamental redesign of the
parallel implementation of the Barnes-Hut gravitational tree algorithm. 
In this method we fully use all the available
computing resources and apply them where they are best suited. The GPUs are
used for the high throughput parallel methods and the CPUs for the irregular
and communication related methods.

\section*{Acknowledgments}

It is our pleasure to thank Jack Wells for arranging direct access on Titan,
Thomas Schulthess for arranging direct access on Piz Daint,
Arthur Trew and Richard Kenwayand for arranging access to 
HECToR,  all via Directors' Discretion time. 
We also thank Mark Harris, Stephen
Jones and Simon Green from NVIDIA for their assistance in optimizing {\tt
Bonsai} for the NVIDIA K20X. Our gratitude also goes to Daniel Caputo and Alex
Rimoldi for their comments and careful proof-reading of the manuscipt 
and Junichi Baba (ELSI, Tokyo Institute of Technology) for his helpful comments
on the scientific results. Part of
our calculations were performed on Hector (Edinburgh), HA-PACS (University of
Tsukuba), ATERUI (National Astronomical Observatory of Japan) and Little Green
Machine (Leiden University).  This work was supported by the Netherlands
Research Council NWO (grants \#639.073.803 [VICI], \#643.000.802 [JB],
\#614.061.608 [AMUSE] and \#612.071.305 [LGM]), by the Netherlands Research
School for Astronomy (NOVA), NAOJ Fellow, MEXT HPCI strategic program and
KAKENHI (under Grant Number 24740115).  This research used resources of the Oak
Ridge Leadership Computing Facility at the Oak Ridge National Laboratory, which
is supported by the Office of Science of the U.S.  Department of Energy under
Contract No.  DE-AC05-00OR22725.
This work was supported by a grant from the Swiss National 
Supercomputing Centre (CSCS) under project ID d24.

\bibliographystyle{IEEEtran}
\bibliography{gbp}

\end{document}